\documentstyle{article}

\begin{document}

\title{Gravitational Phase Transition in Neutron Stars}
\author{R.P. Lano \\
Centre for Theoretical Studies\\
Indian Institute of Science\\
Bangalore - 560 012\\
India}
\date{October 14, 1996}
\maketitle

\begin{abstract}
\baselineskip12pt

The possibility of a gravitational phase transition, especially with respect
to neutron stars is investigated. First, a semiclassical treatment is given,
predicting a gravitational London penetration depth of 12km for neutron
stars. Second, the problem is considered from a Ginzburg-Landau point of
view. A gravitational Meissner effect, a gravitational Aharanov-Bohm type
effect and a gravitational ferromagnetic type phase are predicted. Finally,
a field theoretic consideration predicts a mass term for the graviton below
a certain critical temperature.

\vspace{12pt}

\noindent PACS number(s): 04.60, 04.70, 11.15.E, 97.60.J
\end{abstract}

\newpage\ \baselineskip12pt

\section{Introduction}

In the post-Newtonian approximation, gravity looks like an effective U(1)
theory, closely resembling the theory of electrodynamics. In this work we
consider effects arising from the breaking of this U(1) symmetry. Basically,
we consider the possibility of a macroscopic quantum gravity effect in
analogy to the effect of superconductivity in the electromagnetic theory.

Using a semiclassical approach in close analogy to London's original
approach, we find that for neutron stars, macroscopic quantum gravity
effects seem to become significant. We predict a gravitational Meissner
effect with a gravitational London penetration depth for neutron stars of
about 12km, which is somewhat smaller than the neutron star's radius.

A Ginzburg-Landau treatment leads to the same result. We use the close
analogy between electromagnetism and the post-Newtonian approximation and
predict again the Meissner effect, a gravitational Aharanov-Bohm effect and
a gravitational ferromagnetic type phase, where all the spin angular momenta
of the neutron star become aligned.

Finally we look at the graviton. We use the gravitational wave expansion
from general relativity, treating the graviton as a spin two field. Adding
the Ginzburg-Landau terms through the energy momentum tensor, we observe
that the theory undergoes a phase transition, with the graviton becoming
massive below a certain critical temperature. Again, this predicts a
Meissner effect, justifying the semi-classical treatment.

\section{Motivation}

Neutron stars are extraordinarily fascinating objects: Not only are they our
preferred laboratory to test general relativity, but they also seem to
exhibit macroscopic quantum effects, i.e., it is believed that they are
superfluid and maybe superconducting \cite{Pin-92}. Although these two
issues seem to be separate, we will consider here the possibility of a
macroscopic gravitational quantum effect, in close analogy to the case of
superconductivity.

Neutron stars are made mostly out of neutrons. They have usual radii of
about $15$ km and masses of about $1.4$ times the mass of the sun.
Theoretical calculations indicate that for low enough temperatures ($<10^9$%
K) the neutrons are expected to pair up \cite{Sau-89}, similar to nucleons
forming pairs inside nuclei in the Interacting Boson Model (IBM) \cite
{Iac-93, Muk-89}. The calculations indicate that in the crust region the
singlet $^1S_0$ state should be the preferred one, however, for the majority
the triplet $^3P_2$ state should be the preferred state \cite{Alp-89,
Bay-77, Tam-92}. These neutron pairs are bosons by statistics and bose
condensation is expected. Interesting enough, there is experimental evidence
for the pairing: one, it comes from the theory of cooling of neutron stars,
where the reduction of the specific heat due to extensive pairing is
essential \cite{Pag-94}, and two, evidence for superfluidity can be found in
the glitches in the timing history of pulsars \cite{Pin-92, Sau-89}.

Therefore, we may assume that there exists a bose condensate inside a
neutron star. This bose condensation then is responsible for the breaking of
the translational symmetry leading to superfluidity \cite{Mer-78} and
possibly the electromagnetic U(1) symmetry leading to superconductivity \cite
{Mig-59, Sau-89}. Since gravity in the post-Newtonian approximation is an
effective U(1) theory, we may wonder what would happen if in this
approximation, this gravitational U(1) symmetry were to be broken.

\section{Semiclassical Treatment}

For weak gravitational fields and low velocities, Einstein's field equations
can be written in the post-Newtonian approximation \cite{Mis-73, Wei-72}. We
are mostly interested in magnetic-type gravity, and therefore we shall use
the truncated and rewritten version of the parametrized-post-Newtonian (PPN)
formalism given by Braginsky et. al. \cite{Bar-77}. In this formalism,
Einstein's field equations can be rewritten in a form very much resembling
that of Maxwell's equations

\begin{equation}
{\bf \nabla \cdot g=-}4\pi \rho _0+{\cal O}\left( c^{-2}\right) {\rm {%
,\qquad \ }{\bf \nabla \times g=}-\frac 1c\frac{\partial {\bf H}}{\partial t}%
}  \label{E2.3}
\end{equation}

\begin{equation}
{\bf \nabla \cdot H}=0{\rm {,\qquad \ }{\bf \nabla \times H}=4\left( -4\pi
\rho _0\frac{{\bf v}}c+\frac 1c\frac{\partial {\bf g}}{\partial t}\right) ,}
\label{E2.4}
\end{equation}
where $\rho _0$ is the density of rest mass in the local rest frame of the
matter and ${\bf v}$ is the ordinary velocity of the rest mass relative to
the PPN coordinate frame. The equations of motion of an uncharged particle
are identical to the electromagnetic Lorentz force law \cite{Bar-77, Tho-88},

\begin{equation}
\frac{d{\bf v}}{dt}={\bf g}+\frac 1c\,{\bf v\times H}+{\cal O}\left(
c^{-2}\right) .  \label{E2.5}
\end{equation}

In this weak field, slow motion expansion of general relativity, ${\bf g}$
contains mostly first order corrections to flat space-time, and ${\bf H}$
contains second order corrections. For the solar system, ${\bf g}$ is just
the normal Newtonian gravitational acceleration, whereas ${\bf H}$ is
related to angular momentum interactions and effects due to ${\bf H}$ are
about $10^{12}$ times smaller than those due to ${\bf g}$ \cite{Tho-88}.

With these gravitational Maxwell equations we can derive the equivalent of
London's equations \cite{Lon-48, Lon-50, Schr-64} for gravity. The
derivation is completely analogous to the electromagnetic case \cite{Lan-96}%
. This leads to the second London equation for gravity

\begin{equation}
{\bf \nabla \times j}=-\sqrt{G}\frac{\rho _0}c{\bf H,}  \label{E3.5}
\end{equation}
which in turn predicts the Meissner-Ochsenfeld effect \cite{Mei-33} with a
gravitational London penetration depth of

\begin{equation}
\Lambda _L=\left( \frac{c^2}{16\pi G\rho _0}\right) ^{1/2}=5.18\times
10^{12}\;\rho _0^{-1/2},  \label{E3.8}
\end{equation}
where the density $\rho _0$ is measured in ${\rm kg/m}^3$ and $\Lambda _L$
is given in meters.

Calculating the London penetration depth for neutron stars with approximate
densities of about $\rho _{NS}\cong 2\times 10^{17}{\rm kg/m}^3$, we obtain
a London penetration depth of $12{\rm km}$, which is slightly smaller than
the radius of a neutron star with that density. The London length is
inversely proportional to the density, which means increasing density leads
to decreasing London penetration depth.

As a result, the onset of a gravitational Meissner effect is predicted. In
essence it implies that the gravitational magnetic field caused by the huge
angular momentum of the star, will be expelled from the center of the
neutron star. This could be accomplished through induced
matter-supercurrents in the outer layers of the neutron stars, creating
counter-magnetic fields to expel the gravitational magnetic field from its
interior, in analogy to the electromagnetic case. The gravitational Meissner
effect may lead to significant modifications with respect to the collapse of
a neutron star.

For our derivation of the London equations to be valid we must require that
the London penetration depth $\Lambda _L$ is much larger than the coherence
length $\xi _0$ \cite{deG-66}. A simple upper estimate for the coherence
length yields $\xi _0\leq 10^{-13}{\rm m}$ \cite{Lan-96}, which is larger
than the separation between two nucleons, but much smaller than the London
penetration depth.

\section{Gravitational Ginzburg-Landau Theory}

The previous approach was basically classical in nature. For a
Ginzburg-Landau treatment \cite{Gin-50} of the problem we need a quantum
theory. We first show how to modify Schr\"odinger's equation for the
post-Newtonian approximation. Then we describe the neutron pair bose
condensate. With this in place a standard Ginzburg-Landau treatment leads to
the prediction of Meissner effect and a Aharanov-Bohm type effect. In
addition for the triplet state we predict a macroscopic ordering of the
spins similar to ferromagnetism.

\subsection{Gravitational Schr\"odinger equation}

The first experiment showing the direct influence of gravity on the quantum
mechanical phase of a wave function was done a little more than twenty years
ago by Colella, Overhauser and Werner, now known as the COW-experiment \cite
{Col-75, Rau-86}. They used thermal neutrons in an interferometer type
experiment and showed that rotating the experimental apparatus in the
earth's gravitational field gives an interference pattern exactly as
predicted by the 'gravitational' Schr\"odinger equation (see Sakurai \cite
{Sak-85}), 
\begin{equation}
\left( -\frac{\hbar ^2}{2m}\nabla ^2+m\phi \right) \psi =i\hbar \frac{%
\partial \psi }{\partial t}.  \label{PH-A-1}
\end{equation}

Since then additional experiments have been done, as for example the
influence of the earth's rotation on the phase (Sagnac effect) has been
verified experimentally \cite{Rie-91}. Recently atomic interferometry has
become available, which allows for much higher precision \cite{Bor-96}. In
the near future this should allow for the measurement of such effects as
coupling of a particle's spin to the earth's rotation, as well as second
order correction to the Newtonian potential caused by curvature, and the
somewhat more distant future, Lense-Thirring effect and gravitational waves.

Although these experiments have not been done yet, we are encouraged to
follow the analogy of the electromagnetic treatment and propose the
Schr\"odinger equation for post-Newtonian gravity, 
\begin{equation}
\left( -\frac{\hbar ^2}{2m}\left( {\bf \nabla }-\frac{iq}{\hbar c}{\bf A}%
\right) ^2+q\phi \right) \psi =i\hbar \frac{\partial \psi }{\partial t},
\label{PH-A-2}
\end{equation}
where we identify the minimal coupling 
\begin{equation}
{\bf \nabla \rightarrow \nabla }-\frac{iq}{\hbar c}{\bf A.}  \label{PH-A-3}
\end{equation}
Here, $q=\sqrt{G}m,$ and details for the derivation and units can be found
in the appendix. Please note, that equation (\ref{PH-A-2}) is invariant
under the following gauge transformation: 
\begin{equation}
{\bf A}\rightarrow {\bf A}+{\bf \nabla }\Lambda (x)\qquad \rm{and}\qquad
\psi \rightarrow e^{iq\Lambda (x)/\hbar c}\psi .  \label{PH-A-4}
\end{equation}

One major difference to electrodynamics is that $\left( \phi ,{\bf A}\right) 
$ is not a four-vector, but rather transforms as a particular 0 component of
a rank two tensor, which becomes clear considering its definition in terms
of the metric tensor, 
\begin{equation}
g_{00}=1-2\phi ,\qquad g_{ij}=(1-2\phi )\delta _{ij},\qquad \rm{and}\qquad
g_{0i}=-A_i.  \label{PH-A-5}
\end{equation}
However, in the following construction we never require it to be a four
vector.

\subsection{Long Range Order}

As indicated in the introduction, there is experimental evidence that the
neutrons inside a neutron star are superfluid. This means that there exists
a bose-condensate already, which can be described by a Ginzburg-Landau type
theory. Let us review the basic ideas behind this long-range ordering.

Neutrons are fermions with spin of 1/2. Due to the Pauli exclusion
principle, no one-particle level can have a macroscopic occupation number in
a fermion system, so long range order in the one-particle density matrix is
forbidden. But there can be long range order in the two-particle density
matrix \cite{Mer-78}: 
\begin{equation}
\left\langle r_1s_1,r_2s_2\left| \rho ^{(2)}\right| r_1^{\prime }s_1^{\prime
},r_2^{\prime }s_2^{\prime }\right\rangle =\left\langle \Psi _{s_1}^{\dagger
}(r_1)\Psi _{s_2}^{\dagger }(r_2)\,\Psi _{s_2^{\prime }}(r_2^{\prime })\Psi
_{s_1^{\prime }}(r_1^{\prime })\right\rangle .  \label{PH-C-2}
\end{equation}
In the ordered phase, when the pair of points $r_1,r_2$ is far removed from
the pair $r_1^{\prime },r_2^{\prime },$ this becomes 
\begin{equation}
\left\langle r_1s_1,r_2s_2\left| \rho ^{(2)}\right| r_1^{\prime }s_1^{\prime
},r_2^{\prime }s_2^{\prime }\right\rangle \sim \Phi
_{s_1s_2}^{*}(r_1,r_2)\;\Phi _{s_1^{\prime }s_2^{\prime }}(r_1^{\prime
},r_2^{\prime }),  \label{PH-C-3}
\end{equation}
and such ordering is called pairing.

In general, in a paired system the order parameter $\Phi _{s_1s_2}(r_1,r_2)$
has a richer structure: it is a $2\times 2$ matrix in spin space, each
element of which is a complex function of the difference variable $%
r=r_1-r_2. $ Because the fermion field operators anti-commute it follows
that the order parameter behaves like a two-particle wave function under
interchange of particles $\Phi _{s_1s_2}(r_1,r_2)=-\Phi _{s_2s_1}(r_2,r_1).$
It also follows that the order parameter transforms like a two-particle wave
function under rotations in spin space, rotations in position space, or
Galilean transformations \cite{Mer-78}.

The order parameter can be expanded in a basis of four spin states
consisting of the usual singlet state $\left| 0\right\rangle $, where the
spins of the two neutrons effectively cancel, and three triplet states $%
\left| x_\mu \right\rangle $, where the spins add up to one. The singlet
state is invariant under spin rotations and the triplet states transform
like the x-, y-, and z-components of a vector. We now can write the order
parameter or two-particle wave function as 
\begin{equation}
\Phi _{s_1s_2}(r_1,r_2)=\left\langle s_1,s_2\,|\,\Phi (r_1,r_2)\right\rangle
,  \label{PH-C-7}
\end{equation}
where 
\begin{equation}
\Phi (r_1,r_2)=\phi (r_1,r_2)\left| 0\right\rangle +\psi _\mu
(r_1,r_2)\left| x_\mu \right\rangle  \label{PH-C-8}
\end{equation}
is a combination of singlet $\phi (r_1,r_2)$ and triplet $\psi _\mu
(r_1,r_2) $ order parameters.

\subsection{Singlet $^1S_0$ State}

In the singlet state, the quantum numbers $S$ and $L$ are zero, meaning that
the order parameter $\Phi _{s_1s_2}(r_1,r_2)=\phi (r_1,r_2)$ is a spin
scalar and depends only on the magnitude of $r.$ This is equivalent to the
type of ordering in helium-4, and the only degree of freedom of the order
parameter is associated with the phase degeneracy.

Since only the two-particle wave function $\phi \phi ^{*}$ has a
non-vanishing expectation value, physically measurable properties of a
spatially uniform system can depend only on the product $\phi \phi ^{*}.$
If, in the spirit of Landau, we expand the free energy of ordering in powers
of the order parameter, the first two non-vanishing terms will give a free
energy density of the form 
\begin{equation}
{\cal F}_0=\mu ^2\left| \phi \right| ^2+\lambda \left| \phi \right| ^4.
\label{PH-D-1}
\end{equation}

Slow spatial variations in the order parameter can be treated in the usual
gradient expansion. To leading order there is only one term invariant under
rotations and gauge transformations and ${\cal F}_0$ becomes 
\begin{equation}
{\cal F}=\frac 12K\left| \nabla \phi \right| ^2+\mu ^2\left| \phi \right|
^2+\lambda \left| \phi \right| ^4.  \label{PH-D-2}
\end{equation}

The coupling to gravity can be accomplished through minimal coupling. Recall
equation (\ref{PH-A-3}) and the fact that the mass of the neutron pair is $%
2q,$ we find for the free energy 
\begin{equation}
{\cal F}=\frac K2\left( {\bf \nabla }-i\frac{2q}{\hbar c}{\bf A}\right) \phi
\left( {\bf \nabla }+i\frac{2q}{\hbar c}{\bf A}\right) \phi ^{*}+\mu ^2\phi
\phi ^{*}+\lambda \left( \phi \phi ^{*}\right) ^2.  \label{PH-D-3}
\end{equation}
From here we can obtain the currents associated with the gauge symmetry (\ref
{PH-A-4}) through variation, remembering that the mass of the neutron pair
is $2q.$ Thus we find for the conserved current density 
\begin{equation}
{\bf j}=\frac{qK}\hbar \left( i\phi ^{*}{\bf \nabla }\phi -i\phi {\bf \nabla 
}\phi ^{*}-\frac{4q}{\hbar c}{\bf A}\left| \phi \right| ^2\right) .
\label{PH-D-4}
\end{equation}

Now assume that $\phi =\left| \phi \right| e^{i\varphi }$ has its spatial
variation in the phase $\varphi $ and not in the magnitude $\left| \phi
\right| $, then the current simplifies to 
\begin{equation}
{\bf j}=-\frac{2qK}\hbar \left( {\bf \nabla }\varphi +\frac{2q}{\hbar c}{\bf %
A}\right) \left| \phi \right| ^2.  \label{PH-D-5}
\end{equation}
Since the curl of any gradient vanishes, and since $\left| \phi \right| ^2$
is essentially constant, we immediately deduce the second London equation 
\begin{equation}
{\bf \nabla \times j}=-\frac{4q^2K}{\hbar ^2c}\left| \phi \right| ^2\,{\bf %
\nabla \times A.}  \label{PH-D-6}
\end{equation}

Comparing this with our result from the previous chapter, equation (\ref
{E3.5}), we can use this equation to relate macroscopic to microscopic
quantities, that is, we can identify the density 
\begin{equation}
\rho _0=\frac{4q^2K}{\sqrt{G}\hbar ^2}\left| \phi \right| ^2\,.
\label{PH-D-7}
\end{equation}
Notice that by redefining $\phi \rightarrow \sqrt{K}\phi $ in equation (\ref
{PH-D-2}), we can absorb the $K$ into the $\left| \phi \right| ^2.$
Simultaneously we will also redefine $\mu \rightarrow \sqrt{K}\mu $ and $%
\lambda \rightarrow K^2\lambda .$ This then allows us to express $\left|
\phi \right| ^2$ in terms of known quantities, i.e., 
\begin{equation}
\left| \phi \right| ^2=\frac{\sqrt{G}\hbar ^2}{4q^2}\rho _0.  \label{PH-D-8}
\end{equation}

\subsubsection{Meissner Effect}

One peculiar feature of the Ginzburg-Landau free energy is that its minimum
depends on temperature. For temperatures $T$ above a certain critical
temperature $T_c$, the minimum of the potential will simply be at $\left|
\phi \right| ^2=0.$ When the temperature $T$ is below the critical
temperature $T_c$, i.e., $T<T_c$, then the minimum is at 
\begin{equation}
\left| \phi \right| ^2=-\frac{\mu ^2}{2\lambda }>0.  \label{PH-E-1}
\end{equation}
Furthermore, we assume that $\phi $ varies only very slightly over the
sample, i.e., $\left| \phi \right| ^2$ is essentially constant, 
\begin{equation}
{\bf \nabla \times j}=-\frac{4q^2}{\hbar ^2c}\left| \phi \right| ^2\,{\bf %
\nabla \times A,}  \label{PH-E-2}
\end{equation}
and taking the curl of Ampere's equation (\ref{E2.4}) 
\[
{\bf \nabla \times H}=-\frac{16\pi }c{\bf j} 
\]
we find 
\begin{equation}
\nabla ^2{\bf H}=\pi \left( \frac{8q}{\hbar c}\right) ^2\left| \phi \right|
^2\,{\bf H}=k^2{\bf H,}  \label{PH-E-3}
\end{equation}
where we defined 
\begin{equation}
k^2=\pi \left( \frac{8q}{\hbar c}\right) ^2\left| \phi \right| ^2\,
\label{PH-E-4}
\end{equation}
For a neutron star of density $2\times 10^{17}$kg/m$^3$ we can calculate $k$
using the fact that $k=1/\Lambda _L=7.4\times 10^{-9}1/m^2.$

\subsubsection{Flux Quantization: Gravitational Aharanov-Bohm Effect}

As a result of the Meissner effect, appreciable currents can only flow near
the surface of a gravitational superconductor. As a consequence, if we
integrate the current over a closed ring inside the superconductor, we find
that 
\begin{equation}
0=\oint {\bf j}\cdot {\bf dl}=-\frac{2q}\hbar \left| \phi \right| ^2\oint
\left( {\bf \nabla }\varphi +\frac{2q}{\hbar c}{\bf A}\right) \cdot {\bf dl.}
\label{PH-F-1}
\end{equation}
Then Stokes theorem gives us 
\begin{equation}
\oint {\bf A}\cdot {\bf dl=}\int {\bf \nabla \times A\cdot dS}=\int {\bf %
H\cdot dS}=\Phi _H,  \label{PH-F-2}
\end{equation}
where $\Phi _H$ is the flux enclosed in the ring. Since the order parameter $%
\Phi $ is single-valued, its phase must be an integer multiple of $2\pi $,
and therefore 
\begin{equation}
\oint {\bf \nabla }\varphi \cdot {\bf dl}=2\pi n=\frac{2q}{\hbar c}\Phi _H=%
\frac{2q}{\hbar c}\oint {\bf A}\cdot {\bf dl}  \label{PH-F-4}
\end{equation}
or 
\begin{equation}
\Phi _H=\frac{hc}{2q}n=\Phi _0n  \label{PH-F-5}
\end{equation}
with the quantity $\Phi _0=hc/2q=59.8[m^3/s^2\sqrt{G}]$ being the fluxoid or
flux quantum.

\subsection{Triplet $^3P_2$ State}

Only neutrons in the outer layers of a neutron star will be in the singlet
state. Numerical simulations suggest that most of the neutrons will be found
in the triplet, the $^3P_2$ state \cite{Alp-89, Bay-77, Tam-92}. Therefore,
we will briefly consider the $^3P_2$ state with quantum numbers $S=1,$ $L=1$
and $J=2.$ In the triplet state, $\phi $ vanishes in equation (\ref{PH-C-8})
and the vector $\psi _\mu $ is of the form 
\begin{equation}
\psi _\mu (r)=A_{\mu \nu }\,\widehat{r}_\nu \,\chi \left( \left| r\right|
\right) ,  \label{PH-H-1}
\end{equation}
where $A_{\mu \nu }$ is a complex $3\times 3$ matrix. Under rotations of the
orbital coordinates $A_{\mu \nu }$ transforms as a vector with respect to
index $\nu $, under rotations of the spin coordinates, $A_{\mu \nu }$
transforms as a vector with respect to the index $\mu .$ In general, $A_{\mu
\nu }$ represents all $l=1$ order parameters. To stay in the $^3P_2$
subspace we must restrict the matrix $A$ to be traceless and symmetric. Then
the most general $^3P_2$ Ginzburg-Landau functional to order four, is \cite
{Muz-80, Sau-78, Sau-89, Vul-84} 
\begin{equation}
{\cal F}_0=\frac 13\alpha \;\rm{Tr}AA^{\dagger }+\beta _1\left| \rm{Tr}%
A^2\right| ^2+\beta _2\left( \rm{Tr}AA^{*}\right) ^2+\beta _3\left( \rm{%
Tr}A^2A^{*2}\right) .  \label{PH-H-2}
\end{equation}
This equation represents the most general mean field theory of $^3P_2$
pairing near $T_c$ consistent with the overall rotational and gauge
invariance.

The gradient term for the triplet state looks somewhat more complicated, 
\begin{equation}
{\cal F}_{grad}=\frac{K_1}2\left( \nabla _iA_{\mu \,i}^{*}\right) \left(
\nabla _iA_{\mu \,i}\right) +\frac{K_2}2\left( \nabla _iA_{\mu
\,j}^{*}\right) \left( \nabla _jA_{\mu \,i}\right) +\frac{K_3}2\left( \nabla
_iA_{\mu \,i}^{*}\right) \left( \nabla _jA_{\mu \,j}\right) ,  \label{PH-H-3}
\end{equation}
but it gives a current similar to (\ref{PH-D-4}), predicting a Meissner
effect. Notice, that for example for helium-3 the three coefficients are
nearly the same $K_1=K_2=K_3$ \cite{Mer-78}.

What changes significantly in the triplet state are the minima of the free
energy. Again, above a certain critical temperature, the minimum will be for 
${\cal F}_0=0.$ However, below this temperature there are several different
configurations.

Stability requires that the fourth-order term in the Ginzburg-Landau
functional be positive, which implies that $\beta _2>0,$ and restricts the $%
\beta _1-\beta _3$ plane to the region $\beta _3>-3\beta _2$ and $\beta
_3+2\beta _1>-2\beta _2.$ Minimizing the free energy leads to the following
three subregions \cite{Mer-74, Sau-78}:

Region I: $\beta _3>\left| \beta _1\right| -\beta _1.$ In this region the
order parameter is unique, except for a constant phase factor and a
rotation. At the minimum, the free energy is given through 
\[
{\cal F}_{\min }=-\frac{\alpha ^2}{4\beta _2}. 
\]

Region II: $0>\beta _3>-6\beta _1.$ Except for a constant phase factor and a
rotation the solution is again unique and at the minimum 
\[
{\cal F}_{\min }=-\frac{\alpha ^2}{4\left( \beta _2+\frac 13\beta _3\right) }%
. 
\]

Region III: $\beta _3<-4\beta _1-2\left| \beta _1\right| .$ Here the minimum
is given by any $A$ that is real, traceless and symmetric, and at the
minimum 
\[
{\cal F}_{\min }=-\frac{\alpha ^2}{4\left( \beta _1+\beta _2+\frac 12\beta
_3\right) }. 
\]

\subsubsection{Macroscopic Quantum Coherence}

While the expectation value of the orbital angular momentum along any axis
in region II and III vanishes, for the solution in region I it will be
maximally aligned \cite{Mer-74}. This will lead to an effect analogous to
ferromagnetism or the alignment of spins in helium-3 \cite{Vol-90}. Assuming
all the orbital angular momenta of the $^3P_2$ state are aligned, the
neutron star will have a total, macroscopic spin angular momentum of 
\begin{equation}
S=N\cdot \hbar \sim \frac{M_{NS}}{m_N}\cdot \hbar =1.8\cdot 10^{23}\,\rm{%
Js.}  \label{PH-H-5}
\end{equation}
Comparing this to the approximate $S=2\cdot 10^{40}\,$Js of the Crab pulsar
due to rotation, we find that the pulsar's period would have to slow to a
period of a few hundred million years, for this contribution to become
significant.

Furthermore, one would expect gravitational infinite conductivity to occur.
For example Weinberg shows that no energy gap is necessary, since infinite
conductivity depends only on the spontaneous breakdown of gauge invariance 
\cite{Wei-86}.

\section{The Graviton becomes Massive}

The graviton is a spin two field. This may be the main objection to the
previous section. Here we derive the Meissner effect for a spin two
graviton. We start with the Lagrangian for a scalar field and add a
Ginzburg-Landau potential. After we have the Lagrangian identified, we
proceed to calculate the energy-momentum tensor. Expanding the gravitational
field in the gravitational wave expansion, we find that the graviton becomes
massive below a certain critical temperature, i.e., a phase transition has
occurred and we expect the Meissner effect as a consequence. We then repeat
the calculations for a vector field instead of a scalar, and arrive at very
similar results. We end with a prediction of the graviton's mass.

\subsection{Scalar Field}

Let us consider the Lagrangian of a complex scalar field $\phi $, with terms
at most quadratic in the fields, (and we will leave tadpole like terms out
of consideration). A good choice is found in Veltman \cite{Vel-76} or
Feynman \cite{Fey-95}, and also Callan et. al. \cite{Cal-70}: 
\begin{equation}
{\cal L}=\sqrt{g}\left( \frac 12g^{\mu \nu }\partial _\mu \phi \partial _\nu
\phi ^{*}-\frac 12\left( \mu ^2-2aR\right) \phi \phi ^{*}-\lambda \left(
\phi \phi ^{*}\right) ^2+bR^{\mu \nu }\partial _\mu \phi \partial _\nu \phi
^{*}\right)  \label{MG-A-1}
\end{equation}
We will not consider the $b$ term and also higher terms, since it will imply
more than two derivatives. In addition, we can redefine $\mu ^2\rightarrow
\mu ^2-2aR$ to simplify the above Lagrangian. We find, 
\begin{equation}
{\cal L}=\sqrt{g}\left( \frac 12g^{\alpha \beta }\partial _\alpha \phi
\partial _\beta \phi ^{*}-\frac 12\mu ^2\phi \phi ^{*}-\lambda \left( \phi
\phi ^{*}\right) ^2\right) .  \label{MG-A-2}
\end{equation}

Through variation with respect to the metric we will obtain the
energy-momentum tensor for the above Lagrangian, i.e., 
\begin{equation}
T_{\kappa \lambda }=\frac 1{\sqrt{g}}\frac{\delta {\cal L}}{\delta g^{\kappa
\lambda }},  \label{MG-A-3}
\end{equation}
and we find that the energy momentum tensor is given through 
\begin{equation}
T_{\kappa \lambda }=\frac 12\left[ \partial _\kappa \phi \partial _\lambda
\phi ^{*}-\,g_{\kappa \lambda }\,\left( \frac 12g^{\alpha \beta }\partial
_\alpha \phi \partial _\beta \phi ^{*}-\frac 12\mu ^2\phi \phi ^{*}-\lambda
\left( \phi \phi ^{*}\right) ^2\right) \right] .  \label{MG-A-5}
\end{equation}

We are especially interested in the weak field approximation used for
gravitational waves, i.e., $g^{\mu \nu }=\eta ^{\mu \nu }-\kappa h^{\mu \nu
}.$ Using this expansion, we find for the energy momentum tensor 
\begin{eqnarray}
T_{\kappa \lambda } &=&\frac 12\partial _\kappa \phi \partial _\lambda \phi
^{*}  \label{MG-A-6} \\
&&+\frac 14\left( -\eta _{\kappa \lambda }\,\eta ^{\alpha \beta }+\kappa
\left( \eta _{\kappa \lambda }\,h^{\alpha \beta }-h_{\kappa \lambda }\,\eta
^{\alpha \beta }\right) +\,\kappa ^2h_{\kappa \lambda }\,h^{\alpha \beta
}\right) \partial _\alpha \phi \partial _\beta \phi ^{*}  \nonumber \\
&&+\frac 12\,\eta _{\kappa \lambda }\left( \frac 12\mu ^2\phi \phi
^{*}+\lambda \left( \phi \phi ^{*}\right) ^2\right) +\frac{\,\kappa }%
2h_{\kappa \lambda }\left( \frac 12\mu ^2\phi \phi ^{*}+\lambda \left( \phi
\phi ^{*}\right) ^2\right) .  \nonumber
\end{eqnarray}
Notice that for both possible minimal values of $\left| \phi \right| ^2,$
that is, $\left| \phi \right| ^2=0$ above, and $\left| \phi \right| ^2=-\mu
^2/2\lambda $ below the transition temperature, the vacuum expectation value
of the last two terms vanishes. Next for simplicity neglect terms in $\kappa
^2$ and furthermore assume that all the space-time dependance is in the
phase, i.e., 
\begin{equation}
\partial _\kappa \phi =\partial _\kappa \left( \left| \phi \right|
e^{i\varphi }\right) =i\phi \partial _\kappa \varphi .  \label{MG-A-8}
\end{equation}
We then obtain the following form for the energy momentum tensor 
\begin{equation}
T_{\kappa \lambda }=\frac 12\left[ -\partial _\kappa \varphi \partial
_\lambda \varphi +\frac 12\eta _{\kappa \lambda }\,\partial _\alpha \varphi
\partial ^\alpha \varphi -\frac \kappa 2\eta _{\kappa \lambda }\,h^{\alpha
\beta }\partial _\alpha \varphi \partial _\beta \varphi -\frac \kappa
2h_{\kappa \lambda }\partial _\alpha \varphi \partial ^\alpha \varphi
\right] \left| \phi \right| ^2.  \label{MG-A-9}
\end{equation}

This gives the left hand side of Einstein's equations, 
\begin{equation}
R_{\alpha \beta }+\frac 12\,\eta _{\alpha \beta }R=-8\pi G\;T_{\alpha \beta
}.  \label{MG-A-10}
\end{equation}
For the right hand side we use the Ricci tensor in the gravitational wave
expansion \cite{Wei-72} 
\begin{equation}
R_{\alpha \beta }=\kappa \frac 12\left( \partial _\lambda \partial ^\lambda
h_{\alpha \beta }+\partial _\alpha \partial _\beta h_{\;\lambda }^\lambda
\right) ,  \label{MG-A-11}
\end{equation}
where we made use of the gauge condition $\partial _\lambda h_{\;\alpha
}^\lambda =0$. With this we find the following wave equation for the
graviton 
\begin{eqnarray}
&&\partial _\alpha \partial ^\alpha h_{\kappa \lambda }+\left( \partial
_\kappa \partial _\lambda +\,\eta _{\kappa \lambda }\partial _\mu \partial
^\mu \right) h_{\;\alpha }^\alpha =  \label{MG-A-12} \\
&&-8\pi G\kappa \left[ -\partial _\kappa \varphi \partial _\lambda \varphi
-\frac \kappa 2\eta _{\kappa \lambda }\,h^{\alpha \beta }\partial _\alpha
\varphi \partial _\beta \varphi +\frac 12\left( \eta _{\kappa \lambda
}\,-\kappa h_{\kappa \lambda }\right) \partial _\alpha \varphi \partial
^\alpha \varphi \right] \left| \phi \right| ^2  \nonumber
\end{eqnarray}

Now the vacuum expectation value of $\left| \phi \right| ^2$ above and below
phase transition is given by 
\begin{eqnarray}
\left| \phi \right| ^2 &=&0\qquad \qquad \rm{for}\qquad T>T_C
\label{MG-A-13} \\
\left| \phi \right| ^2 &=&-\frac{\mu ^2}{2\lambda }\qquad \rm{for}\qquad
T<T_C,  \label{MG-A-14}
\end{eqnarray}
and we see that above the critical temperature, the graviton is massless,
but below $T_C$ the graviton has acquired a mass.

Clearly this is a case of spontaneous symmetry breaking. Analogous to the
Higgs mechanism we expect the graviton to acquire three new degrees of
freedom, since it is a spin two field. Those degrees can only come from the
scalar field, which means we need a scalar triplet or three scalars to
conspire together to give one graviton its mass. This seems somewhat awkward
and artificial, but still could be done. However, if the symmetry breaking
field is a vector field instead of a scalar, this scenario becomes much more
natural: One vector field has three degrees of freedom, and hence one
graviton would eat up one vector field to acquire its mass, instead of three
scalar fields. Thus we will consider the vector field next.

\subsection{Vector Field}

We have two reasons to consider the vector field: First, most of the neutron
pairs inside a neutron star are in the triplet state. But second, since the
graviton needs three degrees of freedom to become heavy, it would be more
natural, if the symmetry breaking were done via one vector field instead of
three scalar fields.

The Lagrangian for a spin 1 field $\psi _\alpha $ which has both transversal
and longitudinal degrees of freedom is given through \cite{Buc-92}: 
\begin{equation}
{\cal L}=\sqrt{g}\left( \frac 12D_\alpha \psi _\beta D^\alpha \psi ^\beta
+\frac c2D_\alpha \psi ^\alpha D_\beta \psi ^\beta -\frac 12\left( \mu
^2-2aR\right) \psi _\alpha \psi ^\alpha -\lambda \left( \psi _\alpha \psi
^\alpha \right) ^2\right) ,  \label{MG-B-1}
\end{equation}
where $D_\alpha $ is the covariant derivative, 
\begin{equation}
D_\alpha \psi _\beta =\partial _\alpha \psi _\beta -\Gamma _{\alpha \beta
}^\gamma \,\psi _\gamma .  \label{MG-B-2}
\end{equation}
As before, we will redefine $\mu ^2\rightarrow \mu ^2-2aR$ to simplify the
Lagrangian, 
\begin{equation}
{\cal L}=\sqrt{g}\left( \frac 12D_\alpha \psi _\beta D^\alpha \psi ^\beta
+\frac c2D_\alpha \psi ^\alpha D_\beta \psi ^\beta -\frac 12\mu ^2\psi
_\alpha \psi ^\alpha -\lambda \left( \psi _\alpha \psi ^\alpha \right)
^2\right) .  \label{MG-B-3}
\end{equation}

Again by variation with respect to the metric we obtain the following
energy-momentum tensor 
\begin{eqnarray}
T_{\kappa \lambda } &=&\frac 12\left[ g^{\mu \nu }D_\kappa \psi _\mu
D_\lambda \psi _\nu +g^{\alpha \beta }D_\alpha \psi _\kappa D_\beta \psi
_\lambda -\frac 12\,g_{\kappa \lambda }\,g^{\alpha \beta }g^{\mu \nu
}D_\alpha \psi _\mu D_\beta \psi _\nu \right]  \nonumber \\
&&\ +c\left( D_\kappa \psi _\lambda -\frac 14\,g_{\kappa \lambda }\,D_\beta
\psi ^\beta \right) D_\alpha \psi ^\alpha -\left( \frac 12\mu ^2+2\lambda
\psi _\alpha \psi ^\alpha \right) \psi _\kappa \psi _\lambda  \label{MG-B-11}
\\
&&\ +\frac 12\,g_{\kappa \lambda }\,\left( \frac 12\mu ^2+\lambda \psi
_\alpha \psi ^\alpha \right) \psi _\alpha \psi ^\alpha .  \nonumber
\end{eqnarray}
This looks a little complicated and a few simplifications are in order not
to loose sight of what we are looking for. The first thing to notice is that
the vacuum expectation value of the last term will always be zero, since
above $T_C,$ $\left| \psi \right| ^2=0,$ and below $T_C,$ $\left| \psi
\right| ^2=-\mu ^2/2\lambda .$ Next we observe that the Christoffel symbols
in the gravitational wave expansion \cite{Wei-72}, 
\begin{equation}
\Gamma _{\alpha \beta }^\gamma =\frac 12\eta ^{\gamma \rho }\left( \partial
_\alpha h_{\rho \beta }+\partial _\beta h_{\alpha \rho }-\partial _\rho
h_{\alpha \beta }\right) +{\cal O}\left( h^2\right) ,  \label{MG-B-12}
\end{equation}
are proportional to $\partial ^\gamma h_{\lambda \nu },$ and thus we cannot
expect them to give us a Meissner effect. Therefore, we will concentrate
only on 
\begin{eqnarray}
T_{\kappa \lambda } &=&\frac 12\left[ g^{\mu \nu }\partial _\kappa \psi _\mu
\partial _\lambda \psi _\nu +g^{\alpha \beta }\partial _\alpha \psi _\kappa
\partial _\beta \psi _\lambda -\frac 12\,g_{\kappa \lambda }\,g^{\alpha
\beta }g^{\mu \nu }\partial _\alpha \psi _\mu \partial _\beta \psi _\nu
\right]  \nonumber \\
&&\ +c\left( \partial _\kappa \psi _\lambda -\frac 14\,g_{\kappa \lambda
}\,\partial _\beta \psi ^\beta \right) \partial _\alpha \psi ^\alpha -\left(
\frac 12\mu ^2+2\lambda \psi _\alpha \psi ^\alpha \right) \psi _\kappa \psi
_\lambda .  \label{MG-B-13}
\end{eqnarray}
We expand the metric $g^{\mu \nu }=\eta ^{\mu \nu }-\kappa h^{\mu \nu }$ and
neglect terms in $\kappa ^2$ and $\kappa ^3.$ Furthermore assume that all
the space-time dependance is in the phase, i.e., 
\begin{equation}
\partial _\kappa \psi _\mu =\partial _\kappa \left( \left| \psi _\mu \right|
e^{i\varphi }\right) =i\psi _\mu \partial _\kappa \varphi .  \label{MG-B-15}
\end{equation}
Sorting with respect to $\eta ^{\mu \nu }\psi _\mu \psi _\nu $, we arrive at 
\begin{eqnarray}
\!\!\! &&\!T_{\kappa \lambda }=  \nonumber \\
&&\ \frac 14\left( \left( \left( \eta _{\kappa \lambda }+\kappa h_{\kappa
\lambda }\right) \eta ^{\alpha \beta }-\kappa \eta _{\kappa \lambda
}h^{\alpha \beta }\right) \partial _\alpha \varphi \partial _\beta \varphi
-2\partial _\kappa \varphi \partial _\lambda \varphi \right) \eta ^{\mu \nu
}\psi _\mu \psi _\nu  \nonumber \\
&&\ +\left( \frac 12\left( \kappa h^{\alpha \beta }-\eta ^{\alpha \beta
}\right) \partial _\alpha \varphi \partial _\beta \varphi -\frac 12\mu
^2+2\lambda \left( \kappa h^{\alpha \beta }-\eta ^{\alpha \beta }\right)
\psi _\alpha \psi _\beta \right) \psi _\kappa \psi _\lambda  \label{MG-B-16}
\\
&&\ +\frac \kappa 2\left( \partial _\kappa \varphi \partial _\lambda \varphi
-\frac 12\eta _{\kappa \lambda }\eta ^{\alpha \beta }\partial _\alpha
\varphi \partial _\beta \varphi \right) h^{\mu \nu }\psi _\mu \psi _\nu
+c\left( \kappa h^{\mu \nu }-\eta ^{\mu \nu }\right) \partial _\mu \varphi
\partial _\kappa \varphi \psi _\nu \psi _\lambda  \nonumber \\
&&\ +\frac c4\,\left( \eta ^{\mu \nu }\eta _{\kappa \lambda }\eta ^{\alpha
\beta }-\kappa \left( \eta ^{\mu \nu }\eta _{\kappa \lambda }h^{\alpha \beta
}-\eta ^{\mu \nu }h_{\kappa \lambda }\eta ^{\alpha \beta }+\,\,h^{\mu \nu
}\eta _{\kappa \lambda }\eta ^{\alpha \beta }\right) \right) \partial _\mu
\varphi \partial _\alpha \varphi \psi _\nu \psi _\beta .  \nonumber
\end{eqnarray}

As in the scalar case, the above is going to be on the right-hand side of
Einstein's equations (\ref{MG-A-11}), and thus in the gravitational wave
expansion 
\[
\ \partial _\alpha \partial ^\alpha h_{\kappa \lambda }+\left( \partial
_\kappa \partial _\lambda +\,\eta _{\kappa \lambda }\partial _\mu \partial
^\mu \right) h_{\;\alpha }^\alpha =-8\pi G\;T_{\kappa \lambda }. 
\]

Not unexpected, the energy-momentum tensor for the vector field is a little
more complicated as the one for the scalar field. Direct interaction terms
between the graviton and the vector field like $h^{\mu \nu }\psi _\mu \psi
_\nu $ appear. However, the first line in equation (\ref{MG-B-16}) for the
energy momentum tensor, i.e., 
\[
\frac 14\left( \left( \left( \eta _{\kappa \lambda }+\kappa h_{\kappa
\lambda }\right) \eta ^{\alpha \beta }-\kappa \eta _{\kappa \lambda
}h^{\alpha \beta }\right) \partial _\alpha \varphi \partial _\beta \varphi
-2\partial _\kappa \varphi \partial _\lambda \varphi \right) \eta ^{\mu \nu
}\psi _\mu \psi _\nu , 
\]
will show a phase transition. The vacuum expectation value of this term will
be zero above the transition temperature and non-zero below, since 
\begin{eqnarray}
\left| \psi \right| ^2 &=&0\qquad \qquad \rm{for}\qquad T>T_C
\label{MG-B-17} \\
\left| \psi \right| ^2 &=&-\frac{\mu ^2}{2\lambda }\qquad \rm{for}\qquad
T<T_C.  \label{MG-B-18}
\end{eqnarray}

Although a little more complicated than we may have wanted, the vector field
too will give us a gravitational Meissner effect. This is very comforting in
that it validates the assumptions in the previous sections.

\subsection{Discussion}

As a consequence of the Meissner effect, we expect the graviton to become
massive. We can estimate the mass of the graviton inside a neutron star by
taking the Compton wavelength formula 
\begin{equation}
m_g=\frac \hbar {\lambda _Lc}=2.9\cdot 10^{-47}\rm{kg}=1.6\cdot 10^{-11}%
\rm{eV/c}^2\rm{,}  \label{MG-C-4}
\end{equation}
where $\lambda _L$ is the London penetration depth and we used $\lambda
_L=12 $km. As the graviton acquires a mass we then expect that the
gravitational Coulomb potential is to be replaced locally by a Yukawa type
potential, with possible significant consequences for neutron star collapse,
which are currently under investigation.

\section{Conclusion}

In this paper we tried to shed some light on the problem of breaking the
gravitational symmetry. The approach taken is somewhat of a pedestrian in
nature. As close a path as possible to observable objects and experiments
was taken.

The semiclassical London approach gives us the London penetration length.
The number we obtain for neutron stars, $\lambda _L=12$km, is intriguing and
somewhat unexpected, but is an indication that in neutron stars the breaking
of the gravitational symmetry may play a significant role.

The strength of the Ginzburg-Landau approach lies in that it is based on
laboratory gravitational experiments. The COW experiment was the first one
to show the direct influence of gravity on quantum mechanics. We go one step
further and propose a minimal coupling scheme. From there everything else
falls into place following in close analogy to the electromagnetic case. We
predict again the Meissner effect, a gravitational Aharanov-Bohm effect, and
for the triplet state a gravitational 'ferromagnetic' type phase.

The last section addresses the problem from the gravitational wave expansion
point of view. The spontaneous symmetry breaking character of the phenomenon
becomes apparent. Becoming massive, the graviton will acquire three new
degrees of freedom in the same way as the photon gets a new degree of
freedom in the theory of superconductivity.

\renewcommand{\thesection}{}

\section{Acknowledgment}

I would like to thank Professor V.G.J. Rodgers for starting my interest in
this topic. Furthermore, I wish to thank Prof. Patel for his strong
encouragements and helpful comments, without which this paper would be only
half as long, and I would like to thank Prof. Meurice for some very
enlightening discussions.

\renewcommand{\thesection}{\Alph{section}} \setcounter{section}{0}

\section{Appendix}

\subsection{Remark on Units}

The treatment of the Ginzburg-Landau theory is significantly simplified if
one follows the electromagnetic example and introduces the equivalent of
Gau\ss ian cgs-units. Similarly, as in the electromagnetic case, where the
Coulomb force 
\[
F=\frac 1{4\pi \varepsilon _0}\frac{q^2}{r^2}\qquad \rm{becomes}\qquad F=%
\frac{e^2}{r^2}, 
\]
where the two charges $e$ and $q$ are related through $e=4.80324\cdot
10^{-10}$ esu, one can introduce a new gravitational charge $q,$ and 
\begin{equation}
F=G\frac{m^2}{r^2}\qquad \rm{becomes}\qquad F=\frac{q^2}{r^2},
\label{A.13}
\end{equation}
where, for instance, 
\begin{equation}
q_N=\sqrt{G}m_N=1.35639\cdot 10^{-32}\rm{ gsu}  \label{A.14}
\end{equation}
for a neutron with $m_N=1.66\cdot 10^{-27}$kg. 'gsu' stands for
'gravito-static units' and is gravitational equivalent of 'esu'. Table \ref
{tab1} lists a few units. 

\begin{table}[h] 
\[
\begin{tabular}{|l|l|}
\hline
& units \\ \hline
q = mass charge & gsu = $\sqrt{G}$ kg \\ 
q$^2$ = (mass charge)$^2$ & (gsu)$^2$ = N m$^2$ \\ 
$\rho $ = density & gsu / m$^3$ \\ 
{\bf j }= current density & gsu / m$^2$ s \\ 
{\bf F }= force & N = kg m / s$^2$ \\ 
{\bf g }= electric field & N / gsu \\ 
{\bf H }= magnetic field & N / gsu \\ 
$\phi $ = potential & N m / gsu \\ 
{\bf A }= vector potential & N m / gsu \\ \hline
\end{tabular}
\]
\caption{Gravitostatic units.\label{tab1}}
\end{table}

\subsection{Examples and Numbers}

Any gravitating body will be surrounded by a radial gravitoelectric field 
\begin{equation}
{\bf g}=-\sqrt{G}\frac M{r^2}{\bf e}_r  \label{C.1}
\end{equation}
where the bodies mass $M$ is its gravitoelectric monopole moment.

If the body is rotating, then in addition there will be a gravitomagnetic
dipole field of the form 
\begin{equation}
{\bf H}=\frac{2\sqrt{G}}c\left[ \frac{{\bf S}-3\left( {\bf S}\cdot {\bf e}%
_r\right) {\bf e}_r}{r^3}\right]   \label{C.2}
\end{equation}
where the gravitomagnetic dipole moment is its spin angular momentum ${\bf S}
$ \cite{Tho-88}. Typical numbers for $M$ and $S$ are given in table \ref
{tab2}. 

\begin{table}[t] 
\[
\begin{tabular}{|ccc|}
\hline
\multicolumn{1}{|c|}{} & \multicolumn{1}{c|}{M} & S \\ \hline
\multicolumn{1}{|l|}{Earth} & \multicolumn{1}{c|}{$5.98$$\cdot $10$^{27}$ g}
& $7$$\cdot $$10$$^{40}$ g\thinspace cm$^2$ / s \\ 
\multicolumn{1}{|l|}{Sun} & \multicolumn{1}{c|}{$1.99$$\cdot $10$^{33}$ g} & 
$1.7$$\cdot $$10$$^{48}$ g\thinspace cm$^2$ / s \\ 
\multicolumn{1}{|c|}{Neutron Star} & \multicolumn{1}{c|}{$2.8$$\cdot $10$%
^{33}$ g} & $2$$\cdot $$10$$^{47}$ g\thinspace cm$^2$ / s \\ \hline
\end{tabular}
\]
\caption{Some typical masses and angular momenti.\label{tab2}}
\end{table}

\newpage\ 
\newpage\

\end{document}